# Search for the Lepton Flavour Violating decays $\Upsilon(2S) \to e^{\pm}\mu^{\mp}$ and $\Upsilon(3S) \to e^{\pm}\mu^{\mp}$

Hossain Ahmed[*,1], Nafisa Tasneem[1], and Michael Roney[2] on behalf of the
BABAR collaboration
[1]Saint Francis Xavier University, Antigonish, NS, Canada
[2]University of Victoria, Victoria, BC, Canada
[*]hahmed@stfx.ca

**Abstract:** Charged lepton flavour violating processes are unobservable in the standard model, but they are predicted to be enhanced in several new physics extensions. We present the results of a search for $\Upsilon(2S)$ and $\Upsilon(3S)$ decays to $e^{\pm}\mu^{\mp}$ decays. The search was conducted using data samples consisting of 99 million $\Upsilon(2S)$ and 122 million $\Upsilon(3S)$ mesons, collected at center-of-mass energies of 10.02 and 10.36 GeV, respectively, by the BABAR detector at the SLAC PEP-II $e^+e^-$ collider.

**Introduction**

The observation of neutrino oscillations by the Super-Kamiokande Observatory [1] and the Canadian Sudbury Neutrino Observatories (SNO) [2] indicates Lepton Flavour Violation (LFV) in the neutral lepton sector; however, such an oscillation mechanism cannot induce observable LFV in the charged lepton sector. In the Standard Model (SM), Charged Lepton Flavour Violation (CLFV) is highly suppressed due to the small neutrino masses, e.g., $\left(\frac{\Delta m_\nu^2}{M_W^2}\right)^2 \leq 10^{-48}$[3]. Observation of CLFV is, therefore, a clear sign of new physics (NP) beyond the SM. Searches for electron-tau and muon-tau LFV in decays of bound states of a b quark and b antiquark ($b\bar{b}$) have yielded no evidence of a signal, and upper limits ranging from $2.3 \times 10^{-7}$ to $1.12 \times 10^{-6}$ on their branching fractions have been set [4]. Previous searches for electron-muon LFV in the decay of the $\Upsilon(1S)$ and $\Upsilon(3S)$ $b\bar{b}$ bound states have set 90% CL upper limits of $3.9 \times 10^{-7}$ [5] and $3.6 \times 10^{-7}$[6], respectively. Ref. [6] is the published results for electron-muon LFV in the decay of $\Upsilon(3S)$. In this proceeding paper, we report preliminary results from the first search for electron-muon LFV in the decay of the $\Upsilon(2S)$. Indirect theoretical constraints on LFV decays of vector (i.e., spin = 1, parity = −1) $b\bar{b}$ bound states (referred to as the $\Upsilon(nS)$ mesons, n = 1,2,3,4...) can be derived using an argument based on the non-observation of LFV decays of the muon in conjunction with unitarity considerations [7]. Ref. [7] explained that the most stringent indirect bound on electron-muon LFV decays based on the branching fraction $\mathcal{B}(\mu \to eee) < 1.0 \times 10^{-12}$ can be modified if the virtual $\Upsilon(nS)$ *where* $n = 2,3$ is coupled to $e\mu$ by an amount of $\frac{m_\mu^2}{M_{\Upsilon(nS)}^2}$. We use our result to place constraints on $\frac{\Lambda_{NP}}{g_{NP}^2}$ of New Physics (NP) processes that include LFV, where $g_{NP}$ is the coupling of the NP and $\Lambda_{NP}$ is the energy scale of the NP in a model-independent effective field theory.

Our sample of $\Upsilon(2S)$ and $\Upsilon(3S)$ meson data was collected with the BABAR detector at the PEP-II asymmetric-energy $e^+e^-$ collider at the SLAC National Accelerator Laboratory. The detector was operated from 1999 to 2008 and collected data at the center-of-mass (CM) energies of the $\Upsilon(4S)$ (10.58 GeV), $\Upsilon(3S)$ (10.36 GeV), and $\Upsilon(2S)$ (10.02 GeV) resonances, as well as at energies in the vicinity of these resonances. In this paper, we describe a direct search for LFV decays in a sample of 99 million $\Upsilon(2S)$ decays [8] corresponding to an integrated luminosity of 13.60±0.02 $fb^{-1}$[9] collected during 2008 (referred to as Run 7). Data collected at the $\Upsilon(4S)$ in 2007 (referred to as Run 6) with an integrated luminosity of 78.31±0.35 $fb^{-1}$[9], data taken 40 MeV below the $\Upsilon(4S)$ resonance corresponding to 7.752±0.036 $fb^{-1}$ [9], and data taken 40 MeV below the $\Upsilon(2S)$ resonance corresponding to 2.419±0.017 $fb^{-1}$ [9] constitute control samples. These are used to evaluate non-resonant contributions to the background and to study systematic effects in a signal free sample. We employ a blind analysis strategy [10] in which 0.995 $fb^{-1}$ of the $\Upsilon(2S)$ sample is used solely in the stage prior to unblinding, during which selection criteria are optimized and all systematic uncertainties evaluated. The data sample reserved for the LFV search is based on $(91.6\pm0.92)\times10^6$ $\Upsilon(2S)$ decays, corresponding to 12.61±0.03 $fb^{-1}$, and excludes the 0.995 $fb^{-1}$ sample.

In the BABAR detector, which is described in detail else-where [11, 12], the trajectories of charged particles are measured in a 5-layer silicon vertex tracker (SVT) surrounded by a 40-layer cylindrical drift chamber (DCH). This charged particle tracking system is inside a 1.5 T solenoid with its field running approximately parallel to the $e^+e^-$ beams and together they form a magnetic spectrometer. In order to identify and measure the energies and directional information of photons and electrons, an electromagnetic calorimeter (EMC) composed of an array of 6580 thallium doped CsI crystals, located just inside the superconducting magnet, is used. Muons and neutral hadrons are identified by arrays of resistive plate chambers or limited steamer-tube detectors inserted into gaps in the steel of the Instrumented Flux Return (IFR) [2] of the magnet. The $\Upsilon(4S)$ control sample data for this analysis are restricted to Run 6 to ensure that the control (Run 6) and signal (Run 7) data sets have the same IFR detector configurations following an IFR upgrade program that was completed prior to the beginning of Run 6.

**Sample Selections and Analysis Strategies**

The signature for $\Upsilon(2S) \rightarrow e^\pm\mu^\mp$ events consist of exactly two oppositely charged primary particles, an electron and a muon, each with an energy close to half the total energy of the $e^+e^-$ collision in the CM frame, $E_B$. There are two main sources of background: (i) $e^+e^- \rightarrow \mu^+\mu^-(\gamma)$ events in which one of the muons is misidentified, decays in flight, or generates an electron in a material interaction; and (ii) $e^+e^- \rightarrow e^+e^-(\gamma)$ events in which one of the electrons is misidentified. Background from $e^+e^- \rightarrow \tau^+\tau^- \rightarrow e^\pm\mu^\mp 2\nu 2\bar{\nu}$ is efficiently removed with the kinematic requirements. Two-photon processes and generic $\Upsilon(2S)$ decays to two charged particles with particle misidentification are also potential backgrounds. Event selection proceeds in two stages. In the first stage, a dedicated eμ filter is used to preselect events with only an electron candidate and a muon candidate in the detector. In this filter, all events, in addition to passing either the drift chamber or electromagnetic calorimeter higher-level triggers, are required to have exactly two tracks of opposite charge that are separated by more than $90^0$ in the CM. One of the tracks must pass a very loose electron selection (E/p > 0.8) and the other a very loose muon requirement (E/p < 0.8), where E is the energy deposited in the EMC associated with the track of momentum p. The preselection has an 82% efficiency for signal events, including geometrical acceptance. In the second stage of the analysis, we apply tighter and optimized particle identification (PID) and kinematic criteria. Applying PID to select events with one muon and one electron is the most effective means of reducing the background while maintaining an acceptable efficiency. We report 18.37% signal efficiency and 16 survived $\Upsilon(4S)$ run six events in the blinded luminosity by using the optimized PID selectors.

Kinematic requirements are also applied to suppress $e^+e^- \to \tau^+\tau^- \to e^\pm\mu^\mp 2\nu 2\bar{\nu}$ events, radiative Bhabha and μ-pair events, the $e^+e^- \to e^+e^-e^+e^-$ and $e^+e^- \to e^+e^-\mu^+\mu^-$ two-photon processes, and beam gas interactions. In the $p_e/E_B$ vs $p_\mu/E_B$ plane, where $p_e/E_B$ ($p_\mu/E_B$) is the ratio of the electron (muon) momentum to the beam energy in the CM frame, the distribution of e-μ signal events peaks at (1,1). Events must lie within a circle about that peak: namely we require $\left(\frac{p_e}{E_B} - 1\right)^2 + \left(\frac{p_\mu}{E_B} - 1\right)^2 < 0.01$. The angle between the two lepton tracks in the CM is required to be more than $179^0$. In order to reduce continuum background from μ-pairs and to suppress Bhabha events in which an electron is misidentified as a muon because it passes through the space between crystals, the primary muon candidate is required to deposit at least 50 MeV in the EMC. We require that the lepton tracks fall within the angular acceptance ($24^0 < \theta_{Lab} < 130^0$) of the EMC, where $\theta_{Lab}$ is the polar angle of lepton tracks in the lab frame. Figure 1 shows the eμ invariant mass distribution of the data candidates and background events, as well as the potential signal, after all selection requirements have been applied. No events from the generic(3S) MC sample survive the selection.

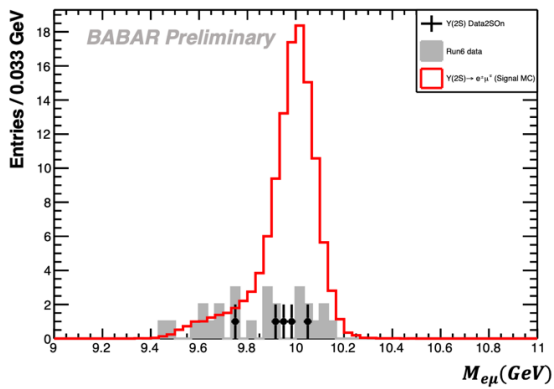

FIG. 1: The distribution of the eμ invariant mass of events surviving all selection criteria. The data sample is presented as the histogram in black with error bars and the open red histogram represents the signal MC with arbitrary normalization. The grey histogram shows the estimate of the continuum background from the Run 6 control sample data with the rate scaled to the amounts expected at 10.023 GeV for a data sample of 12.61 $fb^{-1}$ and the mass scaled to 10.023/10.58.

We assess the systematic uncertainties in the signal efficiency by determining the ratio of data to MC yields for a control sample of $e^+e^- \to \tau^+\tau^- \to e^\pm\mu^\mp 2\nu 2\bar{\nu}$ events in an eμ mass sideband. For this study, we reverse the two major kinematic requirements in the control sample of τ-pair events. The two specific cuts that used to be reversed are the $E_B$-normalized lepton momentum cut and the requirement on the angle between the two tracks. This τ control sample study measures the systematic uncertainty associated with particle identification, tracking, kinematics, trigger selection criteria, and all other quantities except those associated with the two major kinematic requirements used to select the control sample. The signal efficiency is 18.37 ± 0.47%. Figure 2 shows the distribution of $M_{e\mu}$ for the data and MC in the τ control sample.

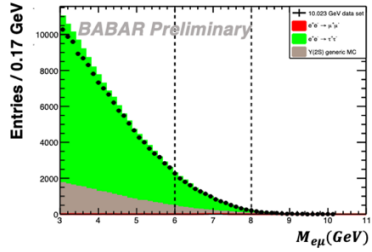

FIG. 2: The distribution of the eμ invariant mass of events in a control sample dominated by τ-pair events obtained by reversing the two major kinematic requirements in the selection.

**Results and Discussions**

After unblinding the data, we find $N_{cand}$ = 5 candidate events and have an expected background of 4.19±0.83 events from a sample of (91.6±0.9)×10⁶ Υ(2S) mesons. Calculating the branching fraction from $(N_{cand}-N_{BG})/(\varepsilon_{sig} N_{\Upsilon(2S)})$ gives:

$$\mathcal{B}(\Upsilon(2S) \to e^\pm\mu^\mp) = (0.5 \pm 1.3(stat) \pm 0.5(syst)) \times 10^{-7} \quad -----(1)$$

where the statistical uncertainty is calculated from $N_{cand}$, and all other uncertainties are included in the systematic uncertainty. As we see no evidence of a signal, we set an upper limit at 90% confidence level (CL) on the branching fraction by using the CLs method [13] including the systematic uncertainties. This is a modified frequentist method that accommodates potential large downward fluctuations in backgrounds. Figure 3 shows the CLs confidence level vs the $\mathcal{B}(\Upsilon(2S) \to e^\pm\mu^\mp)$ and we set a 90% CL upper limit of:

$$\mathcal{B}(\Upsilon(2S) \to e^\pm\mu^\mp) < 3.4 \times 10^{-7} @ 90\% \, CL \quad -----(2)$$

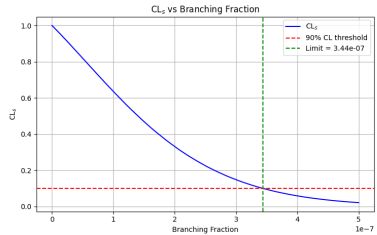

FIG. 3: An upper limit calculation on the branching fraction at 90% confidence level by the CLs method.

This result is the first reported experimental upper limit on $\Upsilon(2S) \to e^{\pm}\mu^{\mp}$. It can be interpreted as a limit on NP using the relationship:

$$\frac{\left(\frac{g_{NP}^2}{\Lambda_{NP}}\right)^2}{\left(\frac{4\pi\alpha_{\Upsilon(2S)}Q_b}{M_{\Upsilon(2S)}}\right)^2} = \frac{\mathcal{B}(\Upsilon(2S) \to e\mu)}{\mathcal{B}(\Upsilon(2S) \to \mu\mu)} \quad ----- (3)$$

and ignoring small kinematic factors. In the relation given in equation 3, $Q_b = -\frac{1}{3}$ is the b-quark charge, $\alpha_{2S}$ is the fine structure constant at the $M_{\Upsilon(2S)}$ energy scale. Using the world average $\mathcal{B}(\Upsilon(2S) \to \mu\mu)$ = 1.93 ±0.17 [5] gives a 90% CL upper limit of $\frac{\Lambda_{NP}}{g_{NP}^2}$ >75 TeV.

We are grateful for the excellent luminosity and machine conditions provided by our PEP-II colleagues, and for the substantial dedicated effort from the computing organizations that support BABAR. The collaborating institutions wish to thank SLAC for its support and kind hospitality. This work is supported by DOE and NSF (USA), NSERC (Canada), CEA and CNRS-IN2P3 (France), BMBF and DFG (Germany), INFN (Italy), FOM (The Netherlands), NFR (Norway), MES (Russia), MINECO (Spain), STFC (United Kingdom), BSF (USA- Israel). Individuals have received support from the Marie Curie EIF (European Union) and the A. P. Sloan Foundation (USA).